\documentclass[10pt,journal,compsoc]{IEEEtran}
\ifCLASSOPTIONcompsoc
  \usepackage[nocompress]{cite}
\else
  \usepackage{cite}
\fi
\usepackage{algorithmic}
\usepackage{graphicx}

\usepackage{textcomp}
\usepackage{hyperref}
\usepackage{cleveref}
\usepackage{hanging}
\usepackage{xcolor}
\usepackage{multirow}
\usepackage{colortbl}
\usepackage{makecell}
\usepackage{boldline}
\usepackage{footmisc}
\usepackage{listings}
\usepackage{tabularx}
\usepackage{setspace}
\usepackage{floatrow}

\definecolor{LightBlue}{rgb}{0.88,1,1}
\definecolor{DarkGreen}{rgb}{0.0,0.6,0.0}
\definecolor{Orange}{rgb}{1.0,0.65,0.0}

\usepackage{hyperref}
\usepackage{cleveref}

\usepackage{amssymb}

\ifCLASSINFOpdf
\else
\fi
\usepackage{array}
\usepackage[hyphenbreaks]{breakurl}
\usepackage{url}

\begin{document}
%
\title{The Relevance of Classic Fuzz Testing:\\Have We Solved This One?}
%
%
%
%

\author{Barton~P.~Miller,~\IEEEmembership{Senior~Member,~IEEE,}
        Mengxiao~Zhang,
        and~Elisa~R.~Heymann
\IEEEcompsocitemizethanks{\IEEEcompsocthanksitem B.P. Miller is with the  
Computer Sciences Department, University of Wisconsin-Madison. 
\protect\\
E-mail: bart@cs.wisc.edu
\IEEEcompsocthanksitem M. Zhang is with the Computer Sciences Department, 
University of Wisconsin-Madison.
\protect\\
E-mail: mzhang464@wisc.edu
\IEEEcompsocthanksitem E. Heymann is with the Computer Sciences Department,
University of Wisconsin-Madison.
\protect\\
E-mail: elisa@cs.wisc.edu}
\thanks{Manuscript received August 13, 2020; revised November 16, 2020.}}

%
%

\markboth{IEEE Transactions on Software Engineering,~Vol.~xx, No.~yy, February~2021, DOI: 10.1109/TSE.2020.3047766}%
{Shell \MakeLowercase{\textit{et al.}}: Bare Demo of IEEEtran.cls for Computer Society Journals}

\newcounter{linuxfailed}
\setcounter{linuxfailed}{9}
\newcounter{linuxtested}
\setcounter{linuxtested}{74}
\newcounter{linuxrate}
\setcounter{linuxrate}{\the\numexpr100*\value{linuxfailed}/\value{linuxtested}}

\newcounter{macosfailed}
\setcounter{macosfailed}{12}
\newcounter{macostested}
\setcounter{macostested}{76}
\newcounter{macosrate}
\setcounter{macosrate}{\the\numexpr100*\value{macosfailed}/\value{macostested}}

\newcounter{freebsdfailed}
\setcounter{freebsdfailed}{15}
\newcounter{freebsdtested}
\setcounter{freebsdtested}{78}
\newcounter{freebsdrate}
\setcounter{freebsdrate}{\the\numexpr100*\value{freebsdfailed}/\value{freebsdtested}}

\newcounter{rustfailed}
\setcounter{rustfailed}{3}
\newcounter{rusttested}
\setcounter{rusttested}{14}
\newcounter{rustrate}
\setcounter{rustrate}{\the\numexpr100*\value{rustfailed}/\value{rusttested}}


\newcounter{totalfailed}
\setcounter{totalfailed}{24}
\newcounter{totaltested}
\setcounter{totaltested}{84}
\newcounter{threeplatformstested}
\setcounter{threeplatformstested}{68}

\IEEEtitleabstractindextext{%
\begin{abstract}
As fuzz testing has passed its 30th anniversary, and in the face of the incredible progress
in fuzz testing techniques and tools, the question arises if the classic, basic fuzz technique
is still useful and applicable?
In that tradition, we have updated the basic fuzz tools and testing scripts and applied them to a large
collection of Unix utilities on Linux, FreeBSD, and MacOS.
As before, our failure criteria was whether the program crashed or hung.
We found that
\arabic{linuxfailed} crash or hang out of \arabic{linuxtested} utilities on Linux,
\arabic{freebsdfailed} out of \arabic{freebsdtested} utilities on FreeBSD,
and
\arabic{macosfailed} out of \arabic{macostested} utilities on MacOS.
A total of
\arabic{totalfailed} different utilities failed across the three platforms.
We note that these failure rates are somewhat higher than our in previous 1995, 2000, and 2006
studies of the reliability of command line utilities.

In the basic fuzz tradition, we debugged each failed utility and categorized
the causes the failures.
Classic categories of failures, such as pointer and array errors and not checking return codes,
were still broadly present in the current results.
In addition, we found a couple of new categories of failures appearing.
We present examples of these failures to illustrate the programming practices that
allowed them to happen.

As a side note, we tested the limited number of utilities available in a modern programming
language (Rust) and found them to be of no better reliability than the standard ones.

\end{abstract}

\begin{IEEEkeywords}
Testing and Debugging;
Testing tools
\end{IEEEkeywords}}

\maketitle

\IEEEdisplaynontitleabstractindextext

%
\IEEEpeerreviewmaketitle

\IEEEraisesectionheading{\section{Introduction}\label{sec:introduction}}

%
%
%
%

\IEEEPARstart{W}{e}
conducted the first fuzz random testing study in the fall of 1988 \cite{bookforward}
and published the first report in 1990~\cite{fuzz1990}.
This early work provided a clear demonstration of the value of simple black-box
random testing by finding real bugs in widely-used software, showing that we could
crash 25-33\% of the utility programs that we tested on each platform.
The simplicity of this technique meant that it was inexpensive and easy to apply.
However, the real contribution lay in the transparent and methodical approach
taken in that study:

\begin{enumerate}
\item
All the tools, test cases, and results were public artifacts that any
software developer or user could download and reproduce the results for themselves
or apply to new software or systems.
\item
We analyzed the results, identifying the source of each crash or hang and
organized these results into categories, trying to come to some understanding
of underlying causes of these bugs in the code.
\item
We socialized and proselytized the techniques and results at all the major Unix
vendors, of the day hoping to entice them into using these techniques or, at least using
the bug reports that we produced to improve their software.
As part of those discussions,
we surveyed our colleagues at the companies and labs that produced versions of
Unix to get insights into their thoughts on these bugs.
\end{enumerate}

Interestingly, the original fuzz tests were occurring at the same time 
as the worm written by Robert Morris Jr. that brought the then-nascent Internet to
its knees \cite{worm1989}.
Our 1990 paper observed that the single largest failure category of the day was
accessing outside the bounds of a buffer and suggested that fuzz testing
might be used "...to help find security holes."
This kind of bug is exactly what Morris exploited in his worm.
And this is still the kind of error that is continuing to cause
massive disruptions in the Internet as evidenced by
the Heart Bleed vulnerability in OpenSSL \cite{openssl}
in 2014 and the more recent
vulnerability in the classic sudo command in 2018~\cite{CVE-2019-18634}.
Compounding this problem is that even the best static analysis tools have had
limited success (if any) in detecting such errors in the code \cite{heartbleed-tools}.

Our initial fuzz testing results were derided and the original paper received a
hostile response from the software engineering conferences and journals of the day;
the reviews were not just negative but outright hostile and insulting.
It took persistence getting that paper published;
there was definitely not an immediately groundswell of adoption of these
techniques.

However, we persisted.
Five years later, our anecdotal experience suggested that little
had changed in software reliability so we repeated and extended these
tests~\cite{fuzz1995}.
And an important new player had arrived on the scene: the open source system.
Even though there were improvements in the reliability of the software from
the commercial Unix vendors, the GNU and Linux utilities showed significantly
lower crash rates than any of these commercial systems.
This was the first concrete evidence that open source software 
was not just the product a bunch of amateurs
pretending to be software developers; they were actually doing as well or
better than the so-called ``professionals''.

The early studies were focused on Unix systems and that led to snide comments by
colleagues from industry that a ``real'' operating system like Microsoft Windows or DEC VMS
would fare better.
Our fuzz study in 2000 of the Windows operating system \cite{fuzz2000} 
and in 2006 of the MacOS operating system \cite{fuzz2006} showed that 
these systems were no more reliable than the Unix variants and, in fact,
noticeably less reliable by our fuzz testing measure.

The 1995 study also started to explore more structured fuzz input.
X Window System  applications communicate with X server via a well-defined message
protocol~\cite{xwindow}.
To test X-Window applications, we first sent unstructured random
input over the connection.
The result was there were some crashes and some cases where the input was rejected,
however the testing reached relatively shallowly into the control flow structure of
the program.
Most of the unstructured input tests stayed in the X Window Protocol processing code.
As a result, we introduced random input that increasingly conformed to random-but-valid user input.
That mean that all fields of the X Protocol messages had
valid values and all mouse events took place in the bounds of a valid window and all
key press events were matched with a corresponding key release event.
As we increased the valid structure on the data, we reached deeper into the control
flow of the utility being tested and found more bugs.

It is important to note that our work on fuzz testing took a simple, perhaps
even simplistic approach to testing.
While it was (and still is) an effective technique in testing and security
assessment, researchers and developers have since made great advances, producing
tools that are both more powerful and more varied in their capabilities.
The past decades have produced a substantial amount of interesting and
useful work in fuzz testing and applying it to many new contexts
\cite{hicks-survey}
\cite{fuzz-survey}
\cite{art-fuzz}
\cite{free-fuzz}.
One recent study
\cite{fuzz-stateoftheart}
reported finding over 170 publications on the topic.

New fuzz tools have taken this type of testing well beyond the simple techniques that
we developed, taking a gray-box approach, allowing them
to dive deeper into a program's control flow
\cite{afl}
\cite{clusterfuzz}
\cite{libfuzzer}
\cite{honggfuzz}
\cite{WhiteSourceBolt}.
However, these more advanced techniques often require more advanced specification of the
input or, for tools that try to find effective erroneous inputs, often require extremely long
execution times as they explore the input and program control-flow space.
If you are a software developer, then it is certainly worth the time to set up these
tools and develop the input strategies that will best test your program.
For studies like ours that survey large bodies of software, our simple 
fuzz tools have the advantage that they are easy to apply and quick to run.

This led us ask the two questions:
First, has operating utility software improved in the way that it handles unexpected input?
Second, are our simple black-box techniques still relevant and are they still
able to find useful errors?
It is now well understood by the software community that reliability is the
foundation of security and that fuzz testing is a powerful first means of
exploration on the path to finding software vulnerabilities.
So we hoped that the kind of flaws found by our original techniques should be
less common, even rare.

Unfortunately, that did not prove to be the case.

In this study, we applied fuzz testing to three Unix variants:


\textbf{Linux:} The most widely used free Unix operating system
of the day.
It is widely used in servers and in the cloud, still appears on desktops and, importantly,
is the foundation of the Android operating system
that appears in billions of mobile devices.
Linux also has the nice characteristic that we studied this system
previously \cite{fuzz1995}.
In our new study,
\arabic{linuxfailed} of the \arabic{linuxtested} utilities that we tested crashed or hung
(\arabic{linuxrate}\%).

\textbf{MacOS:} An extremely popular desktop operating system, which, at its heart,
is based (in part) on FreeBSD.
It is also the foundation of a mobile operating system iOS
that appears in billions of mobile devices.
Again we have the nice characteristic that we studied this system
previously~\cite{fuzz2006}.
In our new study,
\arabic{macosfailed} of the \arabic{macostested} utilities that we tested crashed or hung
(\arabic{macosrate}\%).

\textbf{FreeBSD:} A descendant of the original BSD (Berkeley Software Distribution)
that introduced many key advances to Unix, including the Fast File System \cite{ffs},
the TCP/IP socket programming interface,
and paged virtual memory.
FreeBSD is commonly used in storage servers and appliances.
In our new study,
\arabic{freebsdfailed} of the \arabic{freebsdtested} utilities that we tested crashed or hung
(\arabic{freebsdrate}\%).


As we examined the results from our testing,
we still found too many failures caused by
pointers and arrays being used in loops where the termination condition of the
loop is not properly related to the size of the buffer being processed.
We also saw a noticeable increase in the number of failures where
return values were unchecked or improperly checked.
Interestingly, we saw no cases in this new study where signed characters caused failures.
However, we are seeing the appearance of failures caused by programmers making their
loop conditions and state tracking increasingly (and sometimes incomprehensibly) complex.

Some of the 
failures were caused by bugs that have been present in the
code for years;
regular basic fuzz testing could have caught these much earlier.
Others have been introduced as recently as 2019, such as for \texttt{ftp}.

Of course, we are still dealing with utilities written in C and (increasingly) C++, which
are known for their hazards, especially when dealing with arrays and pointers.
In recent years, system programming languages like Go \cite{go-book} and Rust \cite{rust-book}
have emerged that, among other things, addresses these pointer and array issues.
Some of the ``coreutils'' (a subset of the Unix utilities) have been implemented in these languages,
so it seemed sensible to compare the reliability of these utilities implemented in Go and Rust.
We found a current project implementing the coreutils in 
Rust \cite{rust-repo}.
From this project, we tested the \arabic{rusttested} utilities that
take input and do more than simply copy it.
We also found two projects that were implementing coreutils in Go, but these are quite incomplete
and have not been active for three \cite{go-repo1} and five years \cite{go-repo2}.
Of the \arabic{rusttested} coreutils implemented in Rust that we tested,
\arabic{rustfailed} failed
(\texttt{comm},
\texttt{fold}, and
\texttt{ptx} all hung).
For comparison, one of these \arabic{rusttested} utilities also failed on our Linux,
MacOS and FreeBSD tests: \texttt{ptx} hung on Linux.
Until more utilities, and more complex ones, are reimplemented in these modern languages,
we will have to refrain from making any sweeping statements of comparison.

Section~\ref{sec:fuzz tools} describes how we updated fuzz tools and testing scripts
for current Linux, MacOS, and FreeBSD systems.
Section~\ref{sec:setup} presents our experiments
and
Section~\ref{sec:results} presents our results and the analysis of these results
including assigning the failures to similar categories that we used in earlier studies.
In Section~\ref{sec:discussion}, we wax more philosophical about the testing process
and the implication of these results.
We conclude in Section ~\ref{sec:conclusion}.

\section{Fuzz Tools}
\label{sec:fuzz tools}
The fuzz tools consist of three components:
(1) fuzz, the random input generator,
(2) ptyjig, a tool that can provide input to utilities that use the terminal,
and 
(3) scripts that automate the generation of random input and control the testing of
utilities.
For this study, we updated fuzz and ptyjig, and redesigned the testing scripts.
All source files for the tools and scripts can be found at
\small{\texttt{github.com/dyninst/fuzz}}.

\subsection{Fuzz}
At its heart, the fuzz program is a generator of random characters.
It produces a continuous string of characters on its standard output file,
with variation in the types and amount of characters generated controlled
by the options given to fuzz.
The basic parameter to fuzz specifies the length of generated data in
number of character or number of newline-terminated lines.
The options to fuzz include:

\begin{hangparas}{2em}{1}

\texttt{-p}:
Only the printable ASCII characters.

\texttt{-a}:
The entire 8-bit character set that includes control characters
(it can be important that the high-order bit is set).

\texttt{-0}:
Include the NULL (zero) character in the random data.

\texttt{-l}:
Generate random length strings up to a maximum length specified by the value
associated with the
\texttt{-l} option, with each string terminated by the
newline character.

\texttt{-s}:
Specify the random seed used for input generation.

\texttt{-m}:
Specify a modulus for the random seed, limiting the size
of the seed to the value associated with
\texttt{-m} option minus one.

%
\texttt{-d}:
Specify a delay between each character.
This option is useful when debugging a program to allow the programmer a
chance to observe the utility program's behavior and associate it with the current input.

\end{hangparas}

\noindent
To run fuzz in its simplest form on utility \texttt{as}, you would type:

\indent \texttt{fuzz 1000000 | as}

\noindent
Note that we often use fuzz by first storing the random data in a file and
later running the utility program on that file:

\indent \texttt{fuzz 1000000 > t1; as t1}

\noindent
In our current effort, the fuzz program also was updated to conform to
the more recent
C90~\cite{c90}
and
gnu11~\cite{gnu11}
C standards.

\subsection{Ptyjig}
The purpose of the ptyjig program is to provide input to utilities that read input
from the terminal, a "tty" in Unix parlance or pseudo-tty.
Programs such as \texttt{vim} or \texttt{top} depend on various characteristics of the terminal device, such
as backspace and line-delete functionality and ability to move the cursor around
the terminal's screen.
ptyjig first creates a pseudo-tty device, starts the specified utility, and passes its input
through the pseudo-tty to the utility.
To run fuzz on such a utility, you would type:

\indent \texttt{fuzz 1000000 | ptyjig vim}

Our changes to ptyjig included
(1) updating the arguments to \texttt{wait3},
(2) use the new Posix standard for creating pseudo-tty's,
(3) corrected the way that ptyjig reports the tested utility's completion status,
and
(4) adapted it to work on Linux, MacOS and FreeBSD.

\subsection{Testing Scripts}

The original fuzz distribution included a shell script for controlling the testing
process.
In our current effort,
we developed a new collection of Python scripts to better generate trace data and to
provide more control over how we test the utilities.
One important feature in these scripts is that we automate the use of random input
files that are named on the command line (versus previously assuming that this
random input was only provided through standard input).

Four of the Python scripts
(generate\_small.py,
generate\_medium.py,
generate\_big.py,
and
generate\_huge.py)
generate a broad collection of random files, split along three dimensions:

\begin{enumerate}
\item
Size: The four size categories are: small, medium, large, and huge.
Small files are around 1000 characters long or 10 lines that are up to 100 characters long.
Medium files are around 100,000 characters long or 1000 lines that are up to 100 characters long.
Large files are around 10,000,000 characters long or 100,000 lines that are up to 100 characters long.
Huge files are around 100,000,000 characters long or 1,000,000 lines that are up to 100 characters long.
\item
Characters: There are three character categories: 
all ASCII including the zero byte,
all ASCII excluding the zero byte,
and
only the printable characters.
\item
Newline: The two categories either include the newline character at random intervals
specified by the value associated with \texttt{-l} option or do not treat the newline
character in any special way.
\end{enumerate}
 
\noindent
The fifth Python script, run.py, orchestrates the testing of the utilities.
In other words, it actually runs each specified utility program, setting up the test files
as input.

\indent
\texttt{run.py cnfg\_file -itst\_dir -orslt\_dir}

\noindent
Its basic options are:

\begin{hangparas}{2em}{1}

\texttt{config\_file}: The file that specifies the list of utilities to be run,
the options to pass to that utility, and how the test input will be provided to the utility.

\indent
\texttt{-i}: The directory that contains the input files that will be fed to
each utility program tested.

\texttt{-o}: The directory that will contain the results from each test.
There will be one file for each run of the fuzz tool.

\end{hangparas}

\noindent
Each line in the configuration file is of the form:

\texttt{\{stdin|file|cp|two\_files|pty\} \\
~~~~~~~~~cmd \{cmd\_options\} \{ [ options\_pool ] \}}

\noindent
where 
\texttt{stdin},
\texttt{file},
\texttt{cp},
\texttt{two\_files},
or
\texttt{pty}
control how input is provided to the utility program;
\texttt{cmd} is the name of the utility to run,
\texttt{cmd\_options} are the options (if any) that will be used for each execution of
\texttt{cmd},
and
\texttt{[options\_pool]} is a set of options from which to randomly select
when running the program.
Each option is randomly chosen from the option pool with a probability of 0.5.
For example: 

\noindent
\texttt{stdin bc [-l -w -s -q]}

{\addtolength{\leftskip}{5mm}
Run utility \texttt{bc}, where the random input is provided to standard input.
Randomly select from the pool of options in the square brackets.
}

\noindent
\texttt{file as [-a -D -L -R -v -W -Z -w -x]}

Run utility \texttt{as}, where the input is provided as a file.

\noindent
\texttt{two\_files diff [-s -e -p -T]}

Similar to \texttt{file}, except two input files are provided.

\noindent
\texttt{cp t.c gcc [-c -S -E]}

Run utility \texttt{gcc}, where the random input is first copied to a
file named \texttt{t.c}.
This option is useful for programs like \texttt{gcc} that require
input file with a particular suffix (in this case 
\texttt{".c"}).

\noindent
\texttt{pty vim [-A -b -d]}

Run utility \texttt{vim}, where the random input is fed through a
pseudo-tty.
In addition, interactive utilities often need a specific operation to quit.
For example, in vim, we need to type ESC followed by \texttt{:q!} to quit. 
Therefore, before testing an interactive utility,
the script will append the corresponding end character sequence
to the original test data to prevent a hang that is caused by the lack of an
appropriate quit operation.

The testing script detects when a crash or hang occurs.
To detect a crash, we check the return status of the program.
To detect a hang, we set a timeout argument at 300 seconds.
For each hang result, we verify manually that the program was not just
waiting for more input.

\section{Experimental Set-up}
\label{sec:setup}
\footnotetext[1]{We also ran fuzz tests in 2000 on the Windows operating
systems~\cite{fuzz2000}, but since Windows uses a mostly different set of utilities,
we are not comparing those results in this paper.}
Our testing followed the same basic model as the previous
1990~\cite{fuzz1990},
1995s~\cite{fuzz1995}, and
2006~\cite{fuzz2006}
studies on Unix-based systems using the updated fuzz tools and scripts
described in Section~\ref{sec:fuzz tools}\footnotemark[1].
We evaluated the reliability of up to 80 utility programs on Linux, MacOS and FreeBSD.
The utilities selected needed to take input from standard input or a file, so a utility
like \texttt{ls} is not appropriate.
In addition, the utility needed to do more than trivial processing of the data, so we did not
test utilities like \texttt{cp}.
We also sought utilities based on their availability on more than one platform.

As in our previous studies, we used the primitive criteria that
a program was considered to have failed if it crashed
or hung (stopped responding even though input continued to be available).

Each program was tested on each available platform
using the input files generated by the 
generate\_small.py, generate\_medium.py, generate\_big.py, and generate\_huge.py
scripts.
The random input files generated for this study can be found at
\small{\texttt{ftp://ftp.cs.wisc.edu/paradyn/fuzz/fuzz-2020/}}.

In our previous studies, we tested the utilities
on a variety of Unix platforms, but the majority of them (SunOS, HP-UX, Irix, AIX, Ultrix, NeXT)
are now obsolete.
Our current study focused on three widely-used modern Unix variants:
[1] Linux (Ubuntu 18.04.3 on an Intel 3.20GHz Core i5 using GCC version 7.5.0),
[2] MacOS (Mojave 10.14.5 on an Intel 2.4GHz Core i5 using Apple LLVM version 10.0.0.),
and
[3] FreeBSD (installed on VirtualBox 6.0.14 using the
FreeBSD-11.3-RELEASE-amd64-disc1.iso image and configured with
4GB RAM and 20GB dynamic allocated virtual box disk).

\section{Results}
\label{sec:results}
\floatsetup[table]{capposition=top}
\begin{table*}[t]\centering
\setlength\arrayrulewidth{1pt}
\scriptsize
\centering  
\begin{tabular}{cc}
     \begin{tabular}{|p{1.2cm}<{\centering}|p{1.2cm}<{\centering}|c|p{1.0cm}<{\centering}|c|p{1.0cm}<{\centering}|c|}
     \hline
     \multirow{2}*{Utility}&\multicolumn{2}{c|}{Linux}&\multicolumn{2}{c|}{MacOS}&\multicolumn{2}{c|}{FreeBSD} \\ 
     \cline{2-7}
                  & version& fail      & version    & fail       & version   & fail      \\ 
       \rowcolor{white}
       \hline
       \rowcolor{LightBlue}
       as         & 2.30   & $\circ$    & 11.0.0    &            & {2.17.50} & $\circ$   \\
       \rowcolor{white}
       awk        & 4.1.4  &            & 20070501  &            & 20121220  &           \\
       \rowcolor{LightBlue}
       bash       & 4.4.20 &            & 3.2.57    & $\bullet$  & 5.0.16    &           \\
       \rowcolor{white}
       bc         & 1.07.1 &            & 1.07.1    &            & 1.1       &           \\
       \rowcolor{LightBlue}
       bison      & 3.0.4  & $\circ$    & 3.3       & $\circ$    & 3.4.2     &           \\
       calendar   & *      &            & 1.19      & $\bullet$  & 8.3       &           \\
       \rowcolor{white}
       cat        & 8.28   &            & 1.32      &            & 8.2       &           \\
       \rowcolor{LightBlue}
       checknr    & $-$    &            & 1.9       &            & 8.1       & $\bullet$ \\
       \rowcolor{white}
       clang      & 8.0.0  &            & 11.0.0    &            & 8.0.0     &           \\
       cmp        & 3.6    &            & 2.8.1     &            & 8.3       &           \\
       \rowcolor{LightBlue}
       col        & *      &            & 1.19      &            & 8.5       & $\bullet$ \\
       \rowcolor{white}
       colcrt     & *      &            & 1.18      &            & 8.1       &           \\
       colrm      & *      &            & 1.12      &            & 8.2       &           \\
       comm       & 8.28   &            & 1.21      &            & 8.4       &           \\
       compress   & $-$    &            & 1.23      &            & 8.2       &           \\
       csh        & 20110502-5&         & $-$       &            & $-$       &           \\
       \rowcolor{LightBlue}
       ctags      & 25.2   &            & 5.8\_1    & $\bullet$  & 8.4       & $\bullet$ \\
       \rowcolor{white}
       cut        & 8.28   &            & 1.30      &            & 8.3       &           \\
       dash       & 0.5.10.2-6&         & *         &            & 0.5.10.2  &           \\
       \rowcolor{LightBlue}
       dc         & 1.4.1  & $\circ$    & 1.3       & $\bullet$ $\circ$ & 1.3&           \\
       \rowcolor{white}
       dd         & 8.28   &            & 1.36      &            & 8.5       &           \\
       diff       & 3.6    &            & 2.8.1     &            & 2.8.7     &           \\
       ed         & 1.10   &            & *         &            & 1.5       &           \\
       eqn        & 1.22.3 &            & 1.19.2    &            & 1.19.2    &           \\
       ex/vim     & 8.0    &            & 8.1       &            & 8.1       &           \\
       expand     & 8.28   &            & 1.15      &            & 8.1       &           \\
       \rowcolor{LightBlue}
       flex       & 2.6.4  &            & 2.5.35    &            & 2.5.37    & $\bullet$ \\
       \rowcolor{white}
       fmt        & 8.28   &            & 1.22      &            & 8.1       &           \\
       fold       & 8.28   &            & 1.13      &            & 8.1       &           \\
       \rowcolor{LightBlue}
       ftp        & 0.17-34.1&          & $-$       &            & 8.6       & $\bullet$ \\
       \rowcolor{white}
       gcc        & 7.4.0  &            & $-$       &            & 9.2.0     &           \\
       \rowcolor{LightBlue}
       gdb        & 8.1.0  & $\bullet$  & 8.3.1     & $\bullet$  & 6.1.1     & $\bullet$ \\
       \rowcolor{white}
       gfortran   & 7.4.0  &            & $-$       &            & $-$       &           \\
       grep       & 3.1    &            & 2.5.1     &            & 2.5.1     &           \\
       grn        & $-$    &            & 1.19.2    &            & 1.19.2    &           \\
       \rowcolor{LightBlue}
       groff      & 1.22.3 &            & 1.19.2    & $\circ$    & 1.19.2    & $\circ$   \\
       \rowcolor{white}
       head       & 8.28   &            & 1.20      &            & 8.2       &           \\
       htop       & 2.1.0  &            & 2.2.0     &            & 2.2.0     &           \\
       \rowcolor{LightBlue}
       indent     & $-$    &            & 5.17      & $\bullet$  & 5.17      & $\bullet$ \\
       \rowcolor{white}
       join       & 8.28   &            & 1.2       &            & 8.6       &           \\
       \rowcolor{LightBlue}
       less       & 551    & $\bullet$  & 487       & $\bullet$  & 530       &           \\
       lldb       & $-$    &            & 9.0.1     & $\bullet$  & 8.0.0     & $\bullet$ \\
       \hline
   \end{tabular}
   
   \centering 
     \begin{tabular}{|p{1.2cm}<{\centering}|p{1.0cm}<{\centering}|c|p{1.0cm}<{\centering}|c|p{1.0cm}<{\centering}|c|} 
     \hline
     \multirow{2}*{Utility}&\multicolumn{2}{c|}{Linux}&\multicolumn{2}{c|}{MacOS}&\multicolumn{2}{c|}{FreeBSD} \\ 
     \cline{2-7}
                  & version& fail      & version    & fail       & version   & fail      \\ 
       \hline
       \rowcolor{LightBlue}
       look       & *      &            & 1.18.10   &            & 8.2       & $\circ$   \\
       \rowcolor{white}
       m4         & 1.4.18 &            & 1.4.6     &            & 1.4.18\_1 &           \\
       mail       & *      &            & 8.2       &            & 8.2       &           \\
       \rowcolor{LightBlue}
       make       & 4.1    &            & 3.8.1     &            & 8.3       & $\bullet$ \\
       \rowcolor{white}
       md5/md5sum & 8.28   &            & 1.34      &            & *         &           \\
       mig        & $-$    &            & 116       &            & $-$       &           \\
       more       & 2.31.1 &            & $-$       &            & $-$       &           \\
       neqn       &1.22.3  &            &1.19.2     &            & 1.19.2    &           \\
       nm         &2.30    &            &11.0.0     &            & 3504      &           \\
       \rowcolor{LightBlue}
       pdftex     & 6.2.3  &            & 6.2.3     & $\bullet$  & 6.2.1     &           \\
       \rowcolor{white}
       pic        & 1.22.3 &            & 1.19.2    &            & 1.19.2    &           \\
       pr         & 8.28   &            & 1.18      &            & 8.2       &           \\
       \rowcolor{LightBlue}
       ptx        & 8.28   & $\circ$    & $-$       &            & $-$       &           \\
       \rowcolor{white}
       refer      & $-$    &            & 1.19.2    &            & 1.19.2    &           \\
       rev        & 2.31.1 &            & 1.12      &            & 8.3       &           \\
       sdiff      & 3.6    &            & 2.8.1     &            & 1.36      &           \\
       sed        & 4.4    &            & 1.39      &            & 8.2       &           \\
       \rowcolor{LightBlue}
       sh         & $-$    &            & $-$       &            & 8.6       & $\bullet$ \\
       \rowcolor{white}
       soelim     & 1.22.3 &            &1.19.2     &            & *         &           \\
       sort       &8.28    &            & 2.3       &            & 2.3       &           \\
       \rowcolor{LightBlue}
       spell      & 1.0    & $\circ$    & $-$       &            & $-$       &           \\
       \rowcolor{white}
       split      & 8.28   &            & 1.17      &            & 8.2       &           \\
       strings    & 2.30   &            & *         &            & r3614M    &           \\
       strip      & 2.30   &            & *         &            & r3614M    &           \\
       sum        & 8.28   &            & 1.17      &            & *         &           \\
       tail       & 8.28   &            & 101.40.1  &            & 8.1       &           \\
       tbl        & 1.22.3 &            & 1.19.2    &            & 1.19.2    &           \\
       tcsh       & $-$    &            & 6.21.00   &            & 6.20.00   &           \\
       tee        & 1.22.3 &            & 1.6       &            & 8.1       &           \\
       telnet     & 1.14   &            & 1.16      &            & 8.4       &           \\
       \rowcolor{LightBlue}
       tex        & 6.2.3  & $\bullet$  & 6.2.3     &            & 6.2.1     &           \\
       \rowcolor{white}
       top        & 3.3.12 &            & 125       &            & 3.5beta12 &           \\
       tr         & 8.28   &            & 1.24      &            & 8.2       &           \\
       \rowcolor{LightBlue}
       troff      & 1.22.3 & $\bullet$  & 1.19.2    & $\bullet$  & 1.19.2    & $\bullet$ \\
       \rowcolor{white}
       tsort      & 8.28   &            & 1.13      &            & 8.3       &           \\
       ul         & *      &            & 101.40.1  &            & 8.1       &           \\
       uniq       & 8.28   &            & 101.40.1  &            & 8.3       &           \\
       units      & $-$    &            & *         &            & *         &           \\
       \rowcolor{LightBlue}
       \rowcolor{white}
       wc         & 8.28   &            & 1.21      &            & 8.1       &           \\
       xargs      & 4.7.0  &            & 1.57      &            & 8.1       &           \\
       zic        & 2.27   &            & 8.22      &            & 8.22      &           \\
       \rowcolor{LightBlue}
       zsh        & 5.4.2  &            & 5.7.1     &            & 5.7.1     & $\bullet$ \\
       \hline
   \end{tabular}
   \end{tabular}
   \caption{\textbf{Utilities Tested and the Testing Results} 
    \label{tab:results}
   }
\vspace{0.1cm}
87 utilities were tested on Unix, MacOS, and freeBSD, 67 of which were tested on all three systems.

$\bullet$ = crashed,
$\circ$ = hung,
$-$ = unavailable on that system,
* = version information unavailable.
\end{table*}

\floatsetup[table]{capposition=top}
\begin{table}[tbp] 
\setlength\arrayrulewidth{1pt}
\renewcommand\thetable{2}

   \centering  
    \begin{tabular}{|c|c|c|c|}  
     \hline
     Platform & Linux  & MacOS & FreeBSD \\ 
     \cline{1-4}
      \# tested & \arabic{linuxtested} & \arabic{macostested}  & \arabic{freebsdtested}  \\
      \# failed & \arabic{linuxfailed} & \arabic{macosfailed}  & \arabic{freebsdfailed}  \\
      \% failed & \arabic{linuxrate}\% & \arabic{macosrate}\%  & \arabic{freebsdrate}\%  \\ 
      \hline
   \end{tabular}
   \caption{\textbf{Test Statistics for the 85 Total Utilities Tested} 
   }
   \label{tab:summary}

\end{table}

\floatsetup[table]{capposition=top}
\begin{table}[tbp]  
\setlength\arrayrulewidth{1pt}
\scriptsize
\centering  

\begin{tabular}{|p{1.3cm}<{\centering}|p{1.0cm}<{\centering}|p{1.0cm}<{\centering}|p{1.0cm}<{\centering}|p{1.0cm}<{\centering}|}
\hline
  Utilities                   & Linux 1995                       & MacOS 2006                       & Linux 2020                    & MacOS 2020 \\ 
  \hline 
  as                          &                                  & $\bullet$                        & $\circ$                       & \\

  ctags                       & $\circ$                          &                                  &                               & $\bullet$ \\

  \textcolor{DarkGreen}{ex/vim}&                                 & \textcolor{DarkGreen}{$\blacktriangledown$} &                               & \\

  indent                      & $\bullet$ $\circ$                & $\bullet$                        & $-$                           & $\bullet$ \\

  \textcolor{red}{latex/pdftex}&                                 &                                  &                               & \textcolor{red}{$\blacktriangle$} \\

  nroff                       &                                  & $\bullet$                        &                               & $\circ$ \\

  \textcolor{red}{tex}        &                                  &                                  & \textcolor{red}{$\blacktriangle$}    & \\

  troff                       &                                  & $\bullet$                        & $\bullet$                     & $\bullet$ \\

  \textcolor{DarkGreen}{ul}   & \textcolor{DarkGreen}{$\blacktriangledown$} & \textcolor{DarkGreen}{$\blacktriangledown$} &                               & \\
 \hline
 \end{tabular}
 \caption{\textbf{Comparison of Current and Previous Results} 
 \label{tab:compare}
 }
\vspace{0.1cm}
Comparation of utilities that were tested on
Linux or GNU in 1995~\cite{fuzz1995},
MacOS in 2006~\cite{fuzz2006}, and in this current study, 
where each utility failed in at least one of the studies.
Utilities that
\textcolor{DarkGreen}{failed in previous studies but not in current study are
highlighted in green (\textcolor{DarkGreen}{$\blacktriangledown$})};
utilities that
\textcolor{red}{did not fail in previous studies but failed in the current study
are highlighted in red (\textcolor{red}{$\blacktriangle$})}.

  $\bullet$,$\blacktriangledown$,$\blacktriangle$ = utility crashed, $\circ$ = utility hung, $-$ = utility unavailable on that system.


\end{table}

\floatsetup[table]{capposition=top}
\begin{table}[tbp]  
\setlength\arrayrulewidth{1pt}
\scriptsize
\centering  

\begin{tabular}{|p{0.8cm}<{\centering}|p{0.8cm}<{\centering}|p{0.8cm}<{\centering}|p{0.9cm}<{\centering}|p{0.7cm}<{\centering}|p{0.8cm}<{\centering}|p{0.7cm}<{\centering}|}
  \hline
              & Return Values & Pointers and Arrays & Error Handling & Sub-Process & Complex State & Other \\ 
  \hline 
  as          &               &                     &                &             &               &LF \\
  bash        &M              &                     &                &             &               &   \\
  bison       &               &                     &                &             &LM             &   \\
  calendar    &M              &                     &                &             &               &   \\
  checknr     &               &F                    &                &             &               &   \\
  col         &               &F                    &                &             &               &   \\
  ctags       &               &MF                   &                &             &               &   \\
  dc          &               &M                    &                &             &               &LM \\
  flex        &               &                     &F               &             &               &   \\
  ftp         &F              &                     &                &             &               &   \\
  gdb         &LM             &                     &F               &             &               &   \\
  groff       &               &                     &                &MF           &               &   \\
  indent      &               &MF                   &                &             &               &   \\
  less        &M              &L                    &                &             &               &   \\
  lldb        &MF             &                     &                &             &               &   \\
  look        &               &                     &                &             &F              &   \\
  make        &               &F                    &                &             &               &   \\
  pdftex      &               &                     &M               &             &               &   \\
  ptx         &               &                     &                &             &L              &   \\
  sh          &F              &                     &                &             &               &   \\
  spell       &               &                     &                &L            &               &   \\
  tex         &               &                     &L               &             &               &   \\
  troff       &LMF            &                     &                &             &               &   \\
  zsh         &               &                     &                &             &F              &   \\
  \hline

\end{tabular}

 \caption{\textbf{List of Utilities that Failed, Categorized by Cause, Labeled by Platform} 
 \label{tab:cause}
 }
\vspace{0.1cm}
L = Linux, M = MacOS, F = FreeBSD.
\end{table}


We tested
\arabic{totaltested} utilities across three platforms (Linux, MacOS, and FreeBSD),
where \arabic{threeplatformstested}
of the utilities were available on all three platforms.
In this section, we discuss the quantitative results and then look deeper
into the cause of the crashes we found.

Since Linux and MacOS were tested in our
1995~\cite{fuzz1995} and
2006~\cite{fuzz2006} studies, respectively, we include a comparison
of the current results to those earlier studies.

\subsection{Quantitative Test Results}

As in our previous studies, the experiments described in Section~\ref{sec:setup}
produced a noticeable number of failures,
where a failure is a crash or hang.
Table~\ref{tab:results} lists all utilities tested across three platforms,
and Table~\ref{tab:summary} lists the statistics for the three platforms.
There were \arabic{linuxfailed} failures out of \arabic{linuxtested} utilities test on Linux
(\arabic{linuxrate}\%),
\arabic{macosfailed} out of \arabic{macostested} on MacOS
(\arabic{macosrate}\%),
and
\arabic{freebsdfailed} out of \arabic{freebsdtested} on FreeBSD
(\arabic{freebsdrate}\%).
We caused failures in a total of \arabic{totalfailed} different utilities,
10 of which failed on more than one platform,
and three failed on all three platforms
(gdb and troff).
\arabic{rustfailed} of the \arabic{rusttested} coreutils written in Rust that we tested 
hung (\texttt{comm},
\texttt{fold}, and
\texttt{ptx}), compared to the one that hung on
Linux, MacOS, or FreeBSD (\texttt{ptx}).

Some interesting and well-known utilities appear in the list of failures, including
\texttt{as} (the assembler);
\texttt{bash} and
\texttt{dash} (shells);
\texttt{ctags} (C language cross reference generator);
\texttt{flex} (compiler lexical analyzer generator);
\texttt{ftp} (file transfer utility);
\texttt{gdb} (debugger);
\texttt{pdftex}/\texttt{latex}
and \texttt{tex}
(word processor)\footnotemark[2];
\texttt{make} (build system);
and
\texttt{zsh} (shell).
This collection of key utilities should be enough to get our attention.
\footnotetext[2]{\texttt{pdftex} and \texttt{latex} are aliases to the same executable, while
\texttt{tex} is an independent and slightly different program.
The C code for these programs was generated by web2c from the
original code written in Web (an annotated version of Pascal).}

When we reported the results in our earlier studies,
skeptics were quick to point out that some of
the failed utilities were obscure and of little interest, such as \texttt{ul}, a program
that adds underscores to the text.
However, other utilities that appear archaic are still in common use.
For example, the classic \texttt{nroff} word processing utility (and its GNU
successor based on \texttt{groff}) has been around since
the earliest releases of Unix, but have long been obsoleted by programs such
as LaTex in the 1980's and more modern word processors like Word.
However, manual pages displayed from the
\texttt{man} utility are still processed by \texttt{nroff}.
Some care is needed when disregarding these more obscure utilities because they
may still be in active use in system and production scripts.

Even though the list of utilities in the current study is not the same as those
used in the 1995 study, we can still note the overall crash rates.
For Linux, we have \arabic{linuxrate}\% now compared to 9\% in 1995.
Linux was in its infancy then so had fewer utilities available.
For MacOS, we have \arabic{macosrate}\% now compared to 7\% in 2006.
While it is difficult to over generalize based on studies done so
far apart in time, we can at least say that things have not gotten
better.

Lateral studies across such a large period of time are difficult due to
utilities falling out of favor, such as
\texttt{dbx} being replaced by \texttt{gdb} and \texttt{lldb};
\texttt{yacc} being replaced by \texttt{bison};
and
\texttt{vi} being reimplemented as \texttt{vim}.
In Table~\ref{tab:compare}, we see a list of the utilities that were tested in
all on all of  Linux/GNU in 1995, MacOS in 2006, and in current study on Linux
or MacOS, and that \emph{failed in at least one study}.
As we would hope, a couple of the utilities that failed previously,
did not fail in the current study
(\texttt{ex}/\texttt{vim}
and
\texttt{ul}).
There were also a couple of utilities that previously passed but failed
in the current study,
\texttt{latex}/\texttt{pdftex}
and
\texttt{tex}.
Five of the utilities that failed in previous studies still fail in the current
study, though none for the same reason.

In the current study, we used test cases that were much larger than in our
earlier studies.
In the earlier studies, the largest test input that we used was 100KB.
We note that nine of the utilities we tested required more than that amount to crash:
\texttt{as}       hung    with 8.8 MB on Linux and FreeBSD,
\texttt{calendar} crashed with 4.8 MB on MacOS,
\texttt{checknr}  crashed with 4.8 MB on FreeBSD,
\texttt{col}      crashed with 1   MB on FreeBSD,
\texttt{dc}       hung    with 1.2 MB on Linux and MacOS,
\texttt{flex}     crashed with 4.8 MB on FreeBSD,
\texttt{gdb}      crashed with 9.6 MB on Linux and MacOS,
\texttt{groff}    hung    with 288 KB on MacOS and FreeBSD,
and
\texttt{vgrind}   hung    with 594 KB on FreeBSD.
The remaining utilities crashed with substantially smaller inputs.


\subsection{Examples of Crashes and Hangs}

While the failure statistics are of interest, it is perhaps more important
to understand why these utilities crashed.
For each failed utility, we obtained its source code, debugged the failure,
and then categorized the cause of each failure.
Below is a sample of these results.

\subsubsection{Pointers and Arrays}
\label{sec:pointersandarrays}

From the earliest days, accesses beyond the bound of a buffer in C and C++ have been a problem.
These were the major cause of failure in the first fuzz study and, sadly, are still a major
contributor today.
A loop termination condition should always have an explicit check based
on the size of the buffer and not on its contents.
Of course, it would help to have a language with proper string types
and run-time checking of bounds.

Note that bugs that we describe in this section are distressingly
basic.
We will see more complex bugs in the subsequent sections.

Any unchecked assumptions about the structure of input can be dangerous,
such as we see in
\texttt{\textcolor{blue}{make}}
on FreeBSD.
In function \texttt{Parse\_DoVar} in \texttt{parse.c},
there is a 
for-loop that terminates when an ``='' character is seen and when the
number of right parentheses (or curly brackets) is greater than or
equal to the number of left parentheses (or curly brackets).
\begin{spacing}{1}
\begin{lstlisting}[basicstyle=\ttfamily\small]
  for (depth = 0, cp = line + 1; depth > 0
       || *cp != '='; cp++) {
\end{lstlisting}
\end{spacing}
\noindent
Our test random input had more left parentheses than right, so the loop did
terminate before it went off the end of the buffer.
This bug was introduced somewhat recently, in 2016.

The crash of 
\texttt{\textcolor{blue}{ctags}}
is another simple buffer overflow caused by a loop whose termination condition
does not contain a check on the size of the buffer\footnotemark[3].
\footnotetext[3]{\texttt{GETC(!=,EOF)} is a macro that expands to
\texttt{c = getc(inf) != EOF}.}
\begin{spacing}{1}
\begin{lstlisting}[basicstyle=\ttfamily\small]
  if (xflag)
    for (cp = lbuf; GETC(!=, EOF) && c != '\n';
         *cp++ = c) continue;
  else for (cnt = 0, cp = lbuf; GETC(!=, EOF)
            && cnt < ENDLINE; ++cnt) {
\end{lstlisting}
\end{spacing}
\noindent
Interestingly, this first for-loop lacks the proper check, while the
for-loop that appears in the else-clause a few lines later does
have the proper check.
This bug appears to have been introduced in 1994.

The
\texttt{\textcolor{blue}{indent}}
utility formats a C source code to generate a consistent style of indenting.
When indent is processing a line that has a preprocessor command (starts
with ``\#'') followed immediately by a comment (``/*''), it first finds the
end of the comment and then uses \texttt{bcopy} to copy the characters to
its output buffer.
The loop that finds the comment operates correctly, but in our test case,
finds a comment whose length was longer than that of the output buffer.
The code leading up to the \texttt{bcopy} never checked the length, resulting
in a buffer overwrite.
This bug appear to have been introduced in 1994 and fixed in 2018.

\texttt{\textcolor{blue}{col}}
is a utility that scans text for reverse line feeds caused by the vertical tab character,
\texttt{\^{}K}, and eliminates these characters by reordering the text so that they are
not necessary.
If \texttt{\^{}K} is the first character in the input, there is an error condition that
allows a null pointer to be dereferenced.
The col utility uses a two-pass algorithm to process the text.
On the second pass
(in function \texttt{flush\_lines} in file \texttt{col.c}),
when \texttt{\^{}K} is the first character processed,
a pointer that is used to move backwards has not yet been initialized,
causing the crash.
This bug appears to have been introduced in 2015.

The
\texttt{\textcolor{blue}{checknr}}
utility scans documents written for
\texttt{nroff}
and
\texttt{troff} to find unknown comments
and mismatched opening and closing delimiters.
For example, some macros must come in pairs,
such as 
\texttt{.TS} and \texttt{.TE} (start and end table definition)
and 
\texttt{.EQ} and \texttt{.EN} (start and end equation definition)
When checknr finds an
opening delimiter, it pushes it into a stack (\texttt{stk}) 
and then checks for a match when it finds the corresponding closing delimiter.
However, the stack is of a fixed size (100, based on defined constant \texttt{MAXSTK}),
with no checks to see if an overflow occurs.
Our random input caused this stack to overflow, resulting in a write to an element
beyond the bounds of the \texttt{stk} array.
This bug appears to have been introduced in 1994.

The crash of \texttt{\textcolor{blue}{less}} on Linux is an old fashioned double free.
This crash happens when there is an attempt to overwrite an existing log file
and the user elects not to do so.
The buffer that holds the file name is deleted first
in function \texttt{use\_logfile} in \texttt{edit.c} and later in \texttt{opt\_o} in \texttt{optfunc.c}.
The second free corrupts memory causing a fault on an unrelated memory access (in our case,
it was a call to \texttt{getc}).
This bug appears to have been introduced in 2018 and quite recently.

\subsubsection{Failure to Check Return Value}
\label{sec:returnvalue}

Not checking return value from system calls or functions is a basic
programming mistake.
Often times, this will be an indication of some implicit (rarely documented)
assumptions about parameter values or the behavior of the function being called.

The 
\texttt{\textcolor{blue}{troff}}
text formatter is part of a suite of utilities that share common code.
In code that selects a font, the random input can confuse the selection of
fonts families and styles.
This feature, when combined with trying to embed Postscript in the document will
cause some table look-ups to unexpectedly fail.
The constructor
\texttt{special\_node::special\_node}
calls \texttt{env\_definite\_font}, which has three separate places
in which it checks for an error, causing
\texttt{env\_definite\_font}
to return -1.
However,
\texttt{special\_node::special\_node}
does not check to see if
\texttt{env\_definite\_font}
returned an error:
\lstset{keywordstyle=\color{red},
  morekeywords={foobar},
}
\begin{spacing}{1}
\begin{lstlisting}[basicstyle=\ttfamily\small]
  int fontno = env_definite_font(curenv);
  tf = font_table[fontno]->
   get_tfont(fs,char_height,char_slant,fontno);
\end{lstlisting}
\end{spacing}
\noindent
So, when \texttt{fontno} is -1, it causes an out-of-bounds access in
the next statement when it accesses \texttt{font\_table}.
From the \texttt{gnu.org} git repository,
this bug seems to have been introduced in 2002.

\texttt{\textcolor{blue}{lldb}}
is a newer-generation debugger, built as part of the Clang/LLVM project and written
in C++.
Unfortunately, it has a crash due to an unchecked return value.
When trying to evaluate a command expression, in this case \texttt{`\$0`},
it first looks up the ``Target'', the object indicating which process is being
debugged.
However, for our random input, no valid target was previously defined.

The code in method
\texttt{Target::EvaluateExpression}
in file
\texttt{Target.cpp}
does the look up:
\begin{spacing}{1}
\begin{lstlisting}[basicstyle=\ttfamily\small]
  persistent_var_sp =
     GetScratchTypeSystemForLanguage
       (nullptr, eLanguageTypeC)
       ->GetPersistentExpressionState()
       ->GetVariable(expr);
\end{lstlisting}
\end{spacing}
\noindent
This code is interesting and compact, where the return value from one method
is immediately dereferenced to call another method (and this happens twice).
The problem with such code is that the compactness does not allow for error
checking of the intermediate values.
In this case, our random input causes \texttt{GetScratchTypeSystemForLanguage} to
fail, returning NULL, causing the subsequent dereference to fail.
This bug appears to have been introduced in 1994 and fixed in the repository
in 2019.

\texttt{\textcolor{blue}{ftp}}
is a key network utility that uses the popular
\texttt{libedit}
library to allow line editing of ftp command lines.
Unfortunately, this library is not careful about how it handles
command line parsing errors.
In function
\texttt{tok\_line}
in file
\texttt{tokenizer.c},
if there is an unmatched \textquotesingle\ (single quote) character, then a specific
error value (1) is returned.
In this error case, the parameters (confusingly) named
\texttt{argc} and \texttt{argv} are not initialized.
However, the return value from \texttt{tok\_line} is never checked and
then uninitialized pointer \texttt{argv} is subsequently dereferenced causing a crash.
This bug seemed to have been recently introduced in 2019.

The shell \texttt{\textcolor{blue}{sh}} uses the same \texttt{libedit}
library and will fail under identical input.

\subsubsection{Bad Error Handling}
\label{sec:baderror}

The crash of 
\texttt{\textcolor{blue}{flex}}
is an example of improper error handling (actually, a suppressed error),
which is then followed by a loop with an insufficient termination condition.
Strangely, this is another case of not dealing properly with
matching parentheses.
A variable \texttt{\_sf\_top\_ix} tracks how many elements are on relatively
small stack called
\texttt{\_sf\_stk}.
When a ``('' is read, \texttt{sf\_push} is called (and \texttt{\_sf\_top\_ix} is incremented)
and
when a ``)'' is read, \texttt{sf\_pop}  is called (and \texttt{\_sf\_top\_ix} is decremented).
In our test case, flex reads more right parentheses than left parentheses, calling
\texttt{sf\_pop} to make \texttt{\_sf\_top\_ix} zero, then decremented one more time. 
\texttt{\_sf\_top\_ix} is declared as a \texttt{size\_t}, which is an unsigned long integer
on this platform, so additional decrement of
\texttt{\_sf\_top\_ix}
results in the maximum positive value.

Interestingly, \texttt{sf\_pop} has a check for valid values for\texttt{\_sf\_top\_ix}:
\begin{spacing}{1}
\begin{lstlisting}[basicstyle=\ttfamily\small]
  sf_pop (void)
  {
    assert(_sf_top_ix > 0);
    --_sf_top_ix;
  }
\end{lstlisting}
\end{spacing}
\noindent
There are two major problems with this check.
First, the \texttt{assert} macro had a null definition on the failing platform:
\begin{spacing}{1}
\begin{lstlisting}[basicstyle=\ttfamily\small]
  #ifdef HAVE_ASSERT_H
  #include <assert.h>
  #else
  #define assert(Pred)
  #endif
\end{lstlisting}
\end{spacing}

\noindent
As a result,
\texttt{\_sf\_top\_ix}
becomes zero and the program silently continues.
Next,
\texttt{\_sf\_top\_ix}
is decremented again, this time wrapping to the largest positive number.

Second, even if the check was fixed to be effective, an assert rarely
provides a reasonable or graceful form of error checking.
We note that this bug was introduced in 2011 and
was fixed in a later flex
release but MacOS is still using the buggy version 2.5.35.

The crash of 
\texttt{\textcolor{blue}{pdftex}}/\texttt{\textcolor{blue}{latex}}
is a case of complex processing that never leaves the error handler.
Input was being processed in function
\texttt{getnext}.
Note that the structure of
\texttt{getnext}
is quite complex,
including a switch-statement with 42 cases and
cases with up to nine levels of nested if-statements.
The random input (to standard input) had an invalid character
followed by a sequence of letters.
When
\texttt{getnext} sees an invalid character, it returns and then
\texttt{expand}
is called, which then calls
\texttt{error}
to handle it.
\texttt{error}
then calls
\texttt{gettoken},
which calls
\texttt{getnext},
which calls
\texttt{error} (again, recursively).
\texttt{error}
then calls
\texttt{terminput}
to get another command.
This token happens to be ``Q'', which tells pdftex to go into 
``quiet'' mode where it does not print error messages.
To go into quiet mode, pdftex tries to change the
\texttt{selector}
variable that controls where the output goes by decrementing it.
This is an overly clever way of suppressing printing, as a value of 17 for
\texttt{selector}
means to print to the terminal and 16 means to not print.
Simply setting the variable to 16 would have been less fragile.
Note that values 0--15 mean that output should to one of the file descriptors
stored in the array
\texttt{writefile}.
While still in
\texttt{error},
pdftex calls
\texttt{terminput}
to read the next command and print it.
It does \emph{another} decrement of the
selector variable, which changes it from 16 (no printing) to 15 (printing
to the 15th file descriptor.
Unfortunately, there is no 15th open file, so
\texttt{writefile[15]}
is NULL, causing a dereference of a NULL pointer.
The complexity of the error handling function and general use of huge switch-statements
and highly nested if-statements is certainly a concern for this utility.
This bug appears to have been introduced in 2017.

In the popular
\texttt{\textcolor{blue}{gdb}}
debugger, the crash on FreeBSD provides an example of an
\emph{under-checked} return value.
\texttt{gdb} uses the TUI (Text User Interface) to create windows within the
terminal (shell) window, which is based on the curses library,
This code controls which window to select, but not all the calls to look-up 
functions in this code have their return values checked.
In our test case,
in \texttt{parse\_scrolling\_args}
in \texttt{tui-win.c},
the window look-up fails, returning NULL.

When the window  name is not found, a NULL pointer is returned and
assigned to \texttt{win\_to\_scroll}.
However, the error check only prints out a warning message and allows the program
to continue:
\begin{spacing}{1}
\begin{lstlisting}[basicstyle=\ttfamily\small]
  if (*win_to_scroll == ( . . . ) NULL
      || !(*win_to_scroll)->generic.is_visible)
    warning (" . . .  ");
\end{lstlisting}
\end{spacing}
\noindent
Later, in \texttt{tui\_scroll\_backward}, this pointer is dereferenced causing the crash:
\begin{spacing}{1}
\begin{lstlisting}[basicstyle=\ttfamily\small]
  num_to_scroll=win_to_scroll->generic.height-3;
\end{lstlisting}
\end{spacing}
The bug was introduced with a commit in 1998 and corrected in 2019 in a commit
more recent than the version we had tested.

\subsubsection{Sub-processes}


The text processing utility \texttt{\textcolor{blue}{groff}} hangs because it calls
\texttt{grops} as a sub-process (see 
Section~\ref{sec:complexstate} below).

Another case of vulnerable subprocesses occurs in
\texttt{\textcolor{blue}{spell}}
on Linux.
This utility failed on a relatively long line
(more than 80,000 characters) without any newline character.
In this case, spell creates a subprocess with which it communicates via pipes.
A combination of errors and the long input string causes a deadlock on the pipe
communication.

Spell creates three pipes for parent-child communication:
to send input to the child,
to receive output from the child,
and
to receive error messages from the child.
This is a pretty conventional arrangement to provide stdin, stdout, and stderr
to the child process.

The parent forks a child process that then calls \texttt{execl} to run ispell.
Next, the parent sends input to the child through pipe1 and then waits to read errors or
normal output.
The input is quite long (longer than the pipe's buffer) and has a lot of erroneous data,
so when the child writes to the stderr pipe, it eventually creates a deadlock.
The child sees enough errors to fill up the pipe buffer for stderr, so cannot
continue reading from the stdin pipe;
and the parent is blocked on writing to the stdin pipe, so does not get to the
read on the stderr pipe.
The input line lengths from the parent and quantity of error messages from ispell are both
unexpected behaviors that combine to cause the failure.
This bug appears to have been introduced in 1997.

\subsubsection{Complex State}
\label{sec:complexstate}

In this current study, we caused several utilities to crash or hang, where the 
underlying cause was the tracking of complex state in a loop, often spread
across several functions.
Programmers develop complex criteria in their code that are not well documented nor
symmetrically coded.
As a result, unusual input sequences can cause unexpected patterns of execution.

The \texttt{\textcolor{blue}{zsh}} shell will crash in its line editing mode given an
unexpected sequences of commands.
(Note that this is the third utility that we tested that crashed in code
implementing command line editing. The other two utilities were another shell,
sh, and ftp.)

Our random input triggered the line editing mode with the escape (\texttt{\^{}[}) character.
Once in line editing mode, the next character was a \texttt{d} to indicate a delete
operation, but this character was not followed by a valid qualifier character to indicate
what to delete.
Instead, there was a \texttt{\^{}P} character, indicating that the line editor should
move upward to the previous line of input.
zsh handled this invalid sequence by moving to the previous line but not properly updating
the line length.
Since the previous line was only one character long, the length field now points beyond the
end of a valid string.
The next character of input is a \texttt{j}, which tells the command line editor to
move downward.
This downward scanning code starts by trying to find the end of the current line
in function
\texttt{findeol} in file \texttt{zle\_utils.c}.
However, this scan starts from a point beyond the end of the string.
This bug appears to have been introduced in 1994.

The \texttt{\textcolor{blue}{look}}
utility is a program that looks for lines with a specified string as a prefix;
it will hang when invoked with a file whose first character is the NULL byte.
There are several tests of conditions in the loop in
function \texttt{compare} in \texttt{look.c},
where some of the tests are in four different \texttt{if} statements.
Here, the input byte (the null character)
does not fall into any of the expected categories: alphanumeric character,
end of line, or end of file.
As a result, the pointer to the buffer does not advance.
%
%
This bug appears to have been introduced in 2004.

The well-known parser-generator \texttt{\textcolor{blue}{bison}} will hang on extremely simple
input, where the only character is a carriage return (\texttt{{$\backslash$}r}).
The hang occurs because of a common programming failing: not correctly handling input
lines that do not end with a newline character.
While bison is scanning input character-by-character, if it comes
across this return character, it attempts to generate an error message,
which results in
calling down through six levels of error functions, ending up in
\texttt{location\_caret} in \texttt{location.c}.
In 
\texttt{location\_caret}, there is a loop that
calls \texttt{getc}
in an attempt to scan to the end of the line; this
loop terminates when it finds a newline character.
However, since the last available character has already been read,
the \texttt{getc} function will always return a value of -1. 
The bug was introduced with a commit in 2012 and corrected in 2019 in a commit
more recent than the version we had tested.

The \texttt{\textcolor{blue}{ptx}} indexing utility has a function with
similarly complex state.
In function \texttt{define\_all\_fields},
there is a loop that separates an input line into components.
There is a complex
and inobvious relationship between
the conditions that track which characters to parse and the movement
of a pointer to the current starting point in the buffer.
In this case, one component of the input line, the ``reference'', is longer than
expected,
causing a state where a pointer does not advance, so the loop hangs.
This bug appears to have been introduced in 1998.

In Rust, \texttt{ptx} appeared to hang because of a $N^2$ algorithm in
function \texttt{format\_roff\_line}.
This function processes almost every substring on the input line
that ends with the last character.
Since our input was a single line of 93,290,954 bytes, $N^2$ is $8.7 \times 10^{15}$.
We ran this test case for 34 hours and it finished 22\% of the input, so we
estimate completion time in 155 hours.
While technically not a hang, this ran long enough that we considered it one.

There is a similar story for \texttt{\textcolor{blue}{fold}} in Rust, where
the apparent hang is caused by an algorithm complexity issue.
For an input with $M$ lines and $N$ characters per line, the C version of
\texttt{fold}
has complexity of O($M \times N$), while the Rust version has complexity O($M^2 \times 2$).
The result of this complexity issue is that some of our tests of
\texttt{fold}
will continue for more than 30 hours (the longest we ran a test).

Yet another similar hang occurs in \texttt{\textcolor{blue}{groff}}
when it sends data to the \texttt{\textcolor{blue}{grops}} utility that formats
groff output to PostScript.
The code that sets line width ends up in function \texttt{get\_possibly\_integer\_args} in
file \texttt{ptx.c}.
This function contains a loop that makes up most of the function.
In this loop, state is tracked in a sequence of \texttt{if} statements followed by
a \texttt{switch} statement.
The random input included a \texttt{D} (draw line) command with the
\texttt{t} (thickness) qualifier.
At this point, the code is looking for an integer thickness, but unexpectedly finds
a letter (\texttt{n} in our random input) and reaches a state where it can never
satisfy the loop termination condition.
This bug appears to have been introduced in 2002.

\subsubsection{Others}

Two programs,
\texttt{\textcolor{blue}{dc}}
and
\texttt{\textcolor{blue}{as}}
can be caused to loop for so long that their behavior is practically
indistinguishable from an infinite loop (hang).

\texttt{\textcolor{blue}{dc}}
is a reverse-polish calculator that supports arbitrary-precision arithmetic;
it uses the same arbitrary-length numeric functions as \texttt{bc}.
Our testing input caused an
\emph{apparent} hang, where the
\texttt{bc\_sqrt} function in \texttt{numeric.c} was called
with a parameter that indicated
that the number had 759,375 digits after the decimal point.
After subsequent tests, we saw that this case actually finishes after
six hours (so it might be worth investigating their implementation of the
Newton Raphson Method).

The GNU assembler
\texttt{\textcolor{blue}{as/gas}} is another interesting and strange
example of a utility looping for an extremely long time.
The assembler allows expressions in instructions and directives that have
forward references to symbols defined later in the code.
To resolve forward references, it uses a multi-pass algorithm that stops when it
reaches a fixed point (no symbols that has been used previously changes value).
To prevent an infinite loop when there is a cycle in the expressions,
they limit the loop to execute $N^2$ times, where $N$ is the number of
``fragments'' in the code, where a fragment typically corresponds to a line
of assembly code.
Each time through this loop, the assembler goes through the entire list of
fragments (so this actually has a complexity of $N^3$).

Our random input happened to contain a cycle that is effectively a sequence
like:
\begin{spacing}{1}
\begin{lstlisting}[basicstyle=\ttfamily\small]
    .org  a
    .byte 0
  a:
\end{lstlisting}
\end{spacing}
\noindent
Note that our random input actually had sequence:

\texttt{ .=\~{N} ... JG ... \~{N}:}

\noindent
where the \texttt{.=} is equivalent to \texttt{.org}.
The total size of our random input was quite large, so we estimate that it would take
almost a year before the $N^2$ iterations would be complete.
An algorithm that directly detects cycles in the expression would be
clearer and less prone to unpredictable behaviors.

\section{Discussion}
\label{sec:discussion}
In 1990 fuzz testing was unknown, so finding a lot of bugs that
caused crashes or hangs was not (in retrospect) surprising.
In 1995, well after the results were published and the tool made public,
the failure rate was still significant.
Even at that point in time, many people in the testing community found the technique to
be suspect at best.
Fast forward to 2020, where fuzz testing in its various guises is widely
used by both the testing and security communities,
and where everyday, the tools get smarter and more capable,
we are still seeing failure rates from
\arabic{linuxrate}\% to \arabic{freebsdrate}\%
with the original simple methods.
In 2006, we wrote:
\emph{Plus \c{c}a change, plus c'est la m\^{e}me chose,}
and this still appears to be valid.

In this section, we first discuss some issues that came up during testing
and then present a commentary on the results.

\subsection{Challenges in Bulk Testing}
\label{sec:bulktesting}

The goal of our testing process was to proceed in the most automated way possible.
That means making it easy to set up the tests for a large number of utilities and running
the tests with minimum human intervention.
There were a few issues that made this goal more challenging.

\subsubsection{Avoiding Fatal Side Effects}

Utility programs can sometimes have side effects on the system or other processes.
Some sides, like creating or removing a file, are easy to deal with, while others are more difficult.
For example, in our early fuzz studies,
we learned the obvious-when-you-think-about-it lesson that you cannot
test the \texttt{shutdown} utility.
If you run as a normal user, it does nothing; if you run as root, it takes down the entire system.
More subtly, utilities like \texttt{top} and \texttt{htop} can terminate other processes.
During random testing, we had these programs kill the fuzz tools or important other processes
like the current login shell.
Perhaps testing in a container could help to isolate these affects, something that we plan 
to do in the future.

\subsubsection{Hanging or Just Waiting?}

Some utilities start other programs, and if the utility waits for the new program
to complete, it is difficult to distinguish a hang from a program waiting to be told
to terminate.
Of course shells do this, but there are more subtle cases such as \texttt{less}, which
can start an editor on the file being inspected.

\subsubsection{What You See Is Not Always What You Get}

Another issue that came up is determining exactly what we were testing.
Things are not always what they appear.
For example, the command \texttt{cc} on our Linux platform is a symbolic link to \texttt{gcc}
but on MacOS is a symbolic link to \texttt{clang}.
On MacOS, \texttt{gcc} is an actual binary for \texttt{clang}.
However, if you run \texttt{gcc} on MacOS, you get \texttt{clang} version 11.0.0
while if you explicitly run \texttt{clang}, you get version 9.0.1.
For the \texttt{sh} command on Linux, you get a symbolic link to \texttt{dash},
while the \texttt{sh} command on MacOS is the actual \texttt{bash} binary
and on FreeBSD it is the actual \texttt{sh} binary.
On MacOS, there is also a binary named \texttt{bash}, which is the same
3.2.57(1)-release version as \texttt{sh}.
On MacOS, the \texttt{latex} binary is actually \texttt{pdftex} (a related but not
identical program),
And on MacOS and FreeBSD, the \texttt{more} binary is actually \texttt{less}
(again, a related but not identical program).
The moral is that on any given platform, you need to check carefully as to what
program will run for each command name.

Determining the source of the code and the version number can sometimes be challenging.
For many utilities, you can simply run the program with the
\texttt{-v}
or
\texttt{-}\texttt{-}\texttt{version}
option.
For others, you have to look at the repository or
into the executable file, perhaps using the \texttt{strings} utility to look
for a version string.
Other utilities, such as
\texttt{calendar} and \texttt{col} on Linux, simply defeated our ability to find the version.

\subsubsection{Testing Pseudo-TTY Programs}

A couple of additional issues came up when testing programs meant to run in terminal windows.
First, because an end-of-file indication for the random input does not propagate through the
pseudo-tty, it is difficult to distinguish a program waiting for more input after the end
of random input from a program that is hung.
This has been an issue since the first fuzz testing study.
To avoid this case, for each program tested with ptyjig, we specify a string to append to the
random input to attempt to terminate the utility.
For example, when testing \texttt{vim}, we append the sequence ``ESC \texttt{:} \texttt{q} \texttt{!}''.

Second, if our test input contains certain control sequences such as \texttt{\^{}C},
the testing process can inadvertently kill the utility being tested.
We avoid this case by eliminating such characters from our random input files.

\subsection{Commentary on Results}
\label{sec:commentary}

The utilities that we tested are a dynamic and changing body of code.
So, we expect to see the introduction of new bugs;
no software release is perfect and bug-free.
However, the difference between now and the early fuzz studies is that fuzz testing
is now a widely known and used technique.
And, while there is a definite benefit and attraction from using the newest in fuzz
testing technology, there appear to be some advantages to not discarding the quick and dirty
approach.
In 1990, we wrote: 
\begin{quotation}
\noindent
\emph{Our approach is not a substitute for a formal verification or testing procedure,
but rather an inexpensive mechanism to identify bugs and increase overall system
reliability.}
\end{quotation}
This still appears to be true, both considering basic fuzz testing
to be an easy and effective mechanism and an addition -- not a
replacement -- for other forms of testing.

While many programmers still find
C to be a fun and satisfying language to use, it is well known to be hazardous and has been
widely decried by both the software engineering and security communities.
The continued presence of failures due to misuse of pointers and arrays
(Section~\ref{sec:pointersandarrays}) is evidence that programmers today are
just as vulnerable to these issues as they were 30 years ago.
We need to be using better programming languages.

Some of the world is skeptical.
In an online forum \cite{quora-quote}, we found this strident but pointed commentary:
\begin{quotation}
\noindent
\emph{Oh sure, let's rewrite well tested utilities written in C
in randomly popular language of the month.
You'd spend quite a bit of time rewriting them,
then significantly more time testing them, and then more time
each time you use them because they'd more than likely run slower.
And of course, as they wouldn't be
nearly as mature, you'd regularly encounter bugs and/or exploits
(and yes, languages that aren't C are still susceptible to those).
So after all that is done, you get a grep that is slower, more likely to crash and
overall less likely to work as intended,
but you can at least fill the readme with buzzwords.}
\end{quotation}
Our test results for Rust were not as doom-filled as this poster suggested, but
they were not as good as we hoped.
However, the Rust versions were as fast (or faster) than the standard utilities.
It is interesting to note that all the Rust failures were hangs and not crashes.
These results are limited because
we could not find Rust versions of the more complex utilities.
As a community, we truly need to move away from our legacy C code base,
and give the newly rewritten versions time to mature.
And, we need to continue to test these new versions as they are updated.

The lack of uniform application of well-understood good programming practices is still
problematic.
We still too frequently find
such things as not checking for return values (Section~\ref{sec:returnvalue}) 
and careless error handling (Section~\ref{sec:baderror}).
These are exactly the kinds of errors that we have found in our in-depth software
assessment activities \cite{maritime-journal} and that we teach about in our
Introduction to Software Security course \cite{software-security-course}.
Computer science curriculums need to emphasize these practices at every level 
and in every area, not just in security or software engineering classes.

Two new categories of errors appeared in our current study.
The first new category, which we called ``Complex State'', showed programmers packing more and more
complex conditions and state into a loop.
These loops are often many hundreds of lines long, crossing function boundaries,
with embedded switch-statements
and many-nested if-statements.
Some of this complexity is caused by the incremental accretion of features, without
any effort to clean up, unify, or refactor the code.
Some are just the product of programmers who cannot see the usefulness of simplicity in
their code structure.

The second new category was caused by programs adding line-editing and history
functionality into their utilities.
This occurred in \texttt{ftp}, \texttt{sh}, and \texttt{zsh}.
The authors of the Plan 9 operating system \cite{plan9}, which was considered
the ultimate successor to Unix, factored out this functionality into their window
system.
The result was a single system implementation of the functionality, where individual programs
would not be tempted to reimplement it.

The good news is that several categories of errors that appeared in previous studies did not appear
in our current one.
These are crashes related to end-of-file handling, divide-by-zero, signed characters, and dangerous
input functions.
So, the word has gotten out on some of these problematic programming practices.

We also note that the Unix world is blessed by too much of a good thing.
As you look at Table~\ref{tab:results}, you can see that we often found a different
version of a utility on each platform that we tested.
And utilities that are maintained on one version of Unix are often independent from
those on other versions.
It would be helpful if there was a common code base for each utility, however that
is complicated by differences in operating system library and system call
interfaces.
It is also complicated by changes in language version that may be adopted at different times
by different operating systems (or even different distributions of the same operating system).

Some of the errors that we found have been present in the code for many years, as far
back as 1994 for
\texttt{checknr},
\texttt{ctags},
\texttt{dc}, and
\texttt{indent};
1997 for 
\texttt{spell}; and
1998 for 
\texttt{gdb} and
\texttt{ptx}.
More frequent application of the basic fuzz tests could help to avoid this situation.
And a few errors had been fixed in newer versions of the software than we tested.
Putting together a operating system release is a complex process, and there can
be delays in bringing the latest version of utility into a release.
As with the multiplicity of versions, delays can be influenced by changes in language
versions and changes in operating system library and system call interfaces.

\section{Conclusion}
\label{sec:conclusion}
After more than thirty years, it appears that there is still a place for this type of basic fuzz
testing.
Standard operating system utilities are still crashing at a noticeable rate and not getting better
over time.
And errors seem to persist in the code bases for too long.
If this testing was integrated to the release process of these operating systems, most of
these failures could have been avoided.
Such testing should be paired with careful functional testing and modern fuzz tools.

The prevalence of errors based on pointers and arrays is not surprising.
As long as we use languages like C with inherently unsafe constructs, it will be difficult to
eliminate these errors.
However, as we have found in our security studies, and we can see in this current fuzz study,
there are other categories of errors that can happen in any language.
Good design, good education, ongoing training, testing integrated into the development
cycle, and (perhaps the most important)
a culture that promotes and rewards reliability are all essential to making real progress here.


%


\ifCLASSOPTIONcompsoc
  \section*{Acknowledgments}
\else
  \section*{Acknowledgment}
\fi

\label{sec:ack}
The pioneering fuzz studies done at the University of Wisconsin-Madison
were all conducted as semester projects in Miller's graduate Advanced
Operating Systems (CS736) course.
The work of the students who conducted those studies went well
beyond the reaches of their classroom and continues to influence
software testing and security practices to this day.
These students were:
Lars Fredriksen and Bryan So (1990);
David Koski, Cjin Pheow Lee, Vivekananda Maganty,
Ravi Murthy, Ajitkumar Nataranjan, and Jeff Steidl (1995);
Justin Forrester (2000);
and
Gregory Cooksey and Frederick Moore (2006).

In this study,
we would like to thank Yi (Emma) He for her efforts in helping to revise
the fuzz tools and run the initial experiments.
We also thank James Kupsch and Josef (Bolo) Burger for helping to
find an obscure bug in ptyjig.

\ifCLASSOPTIONcaptionsoff
  \newpage
\fi



\bibliographystyle{abbrv}
\bibliography{references}
%



%

\begin{IEEEbiography}[{\includegraphics[width=1in,height=1.25in,clip,keepaspectratio]
{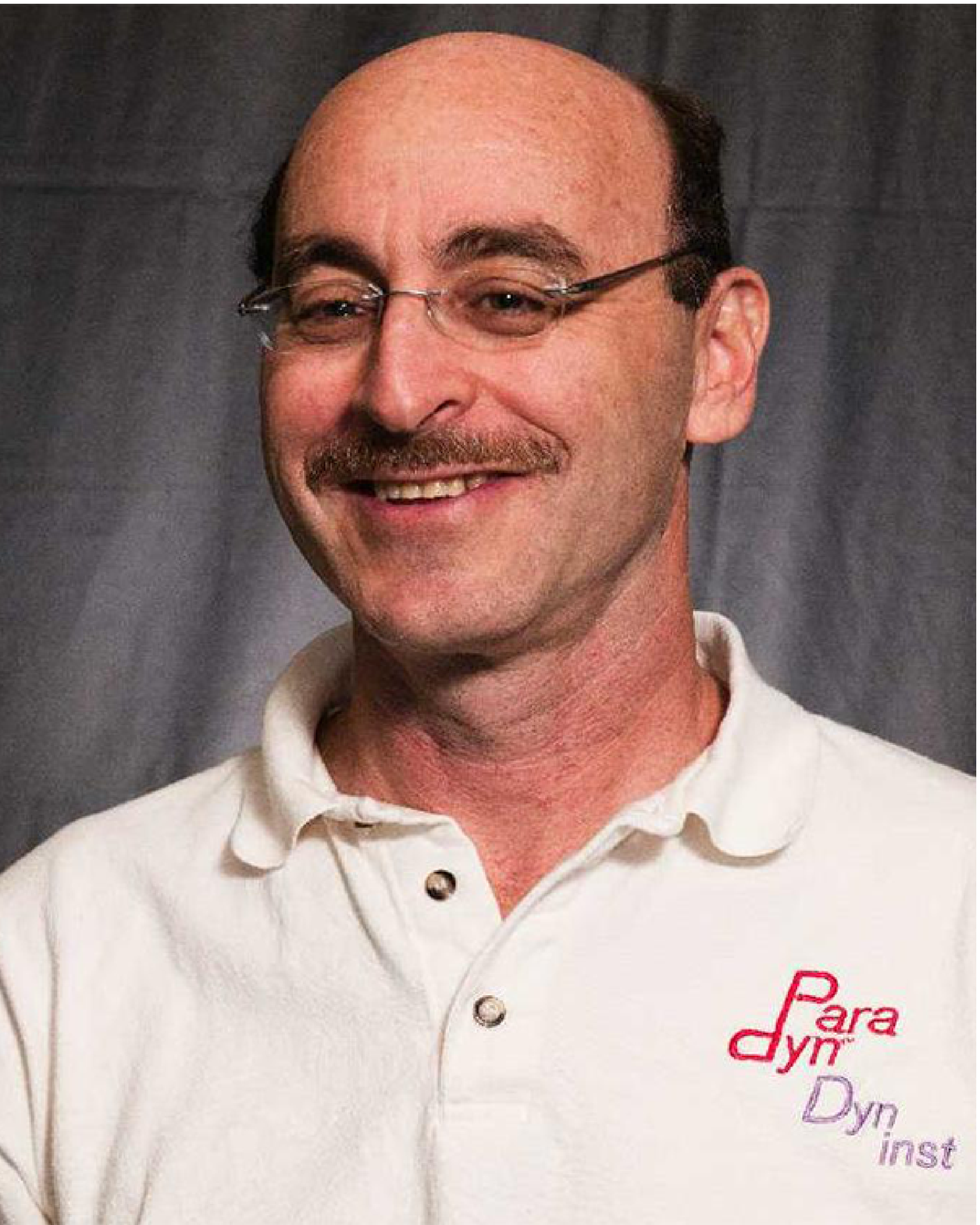}}]{Barton P. Miller}
is the Vilas Distinguished Achievement Professor and Amar \& Belinder Sohi Professor
of Computer Sciences at the University of Wisconsin-Madison.
From 2013-2020, he was Chief Scientist
for the DHS Software Assurance Marketplace research facility and is Software
Assurance Lead on the NSF Cybersecurity Center of Excellence. In addition, he
co-directs the MIST software vulnerability assessment project in collaboration
with his colleagues at the Autonomous University of Barcelona. He also leads the
Paradyn Parallel Performance Tool project, which is investigating performance and
instrumentation technologies for parallel and distributed applications and systems.
His research interests include systems security, binary and malicious code analysis
and instrumentation of extreme scale systems, and parallel and distributed program
measurement and debugging. Miller's research is supported
by the U.S. Department of Energy, National
Science Foundation, NATO, and various corporations. In 1988, Miller founded the
field of fuzz random software testing, which is the foundation of many security
and software engineering disciplines. In 1992, Miller (working with his then-student,
Prof. Jeffrey Hollingsworth), founded the field of dynamic binary code
instrumentation and coined the term ``dynamic instrumentation''. Dynamic
instrumentation forms the basis for his current efforts in malware analysis
and instrumentation. He is a Fellow of the ACM.
\end{IEEEbiography}

\begin{IEEEbiography}[{\includegraphics[width=1in,height=1.25in,clip,keepaspectratio]
{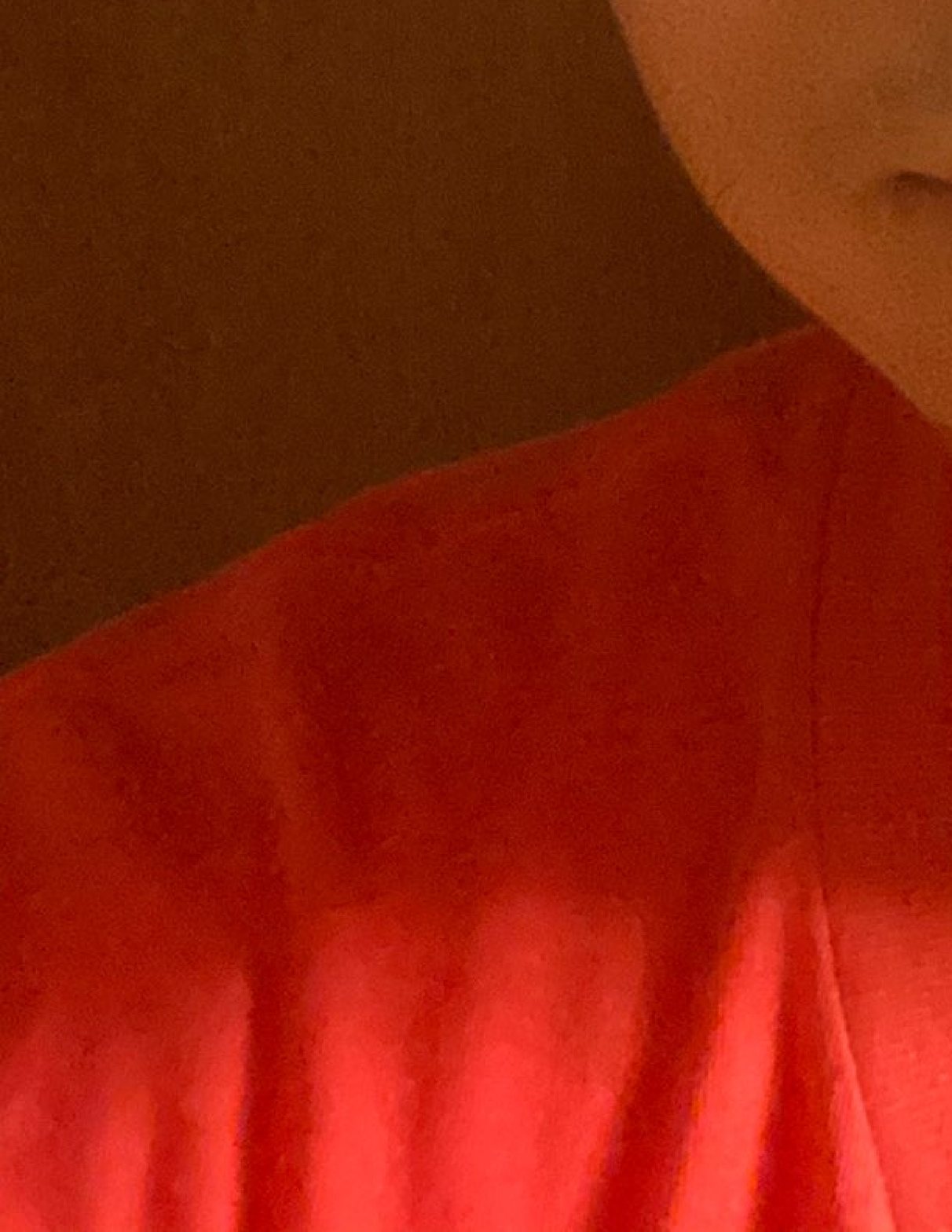}}]{Mengxiao Zhang}
is a masters degree student in Computer Sciences at the University of Wisconsin-Madison.
He earned his bachelor's degree in Computer Science and Engineering from
the Huazhong University of Science and Technology in 2019.
During the masters study, he assisted in research related to software
vulnerability assessment.
He is now a developer of ImageJ, an open platform for scientific image analysis.
\end{IEEEbiography}

\newpage

\begin{IEEEbiography}[{\includegraphics[width=1in,height=1.25in,clip,keepaspectratio]
{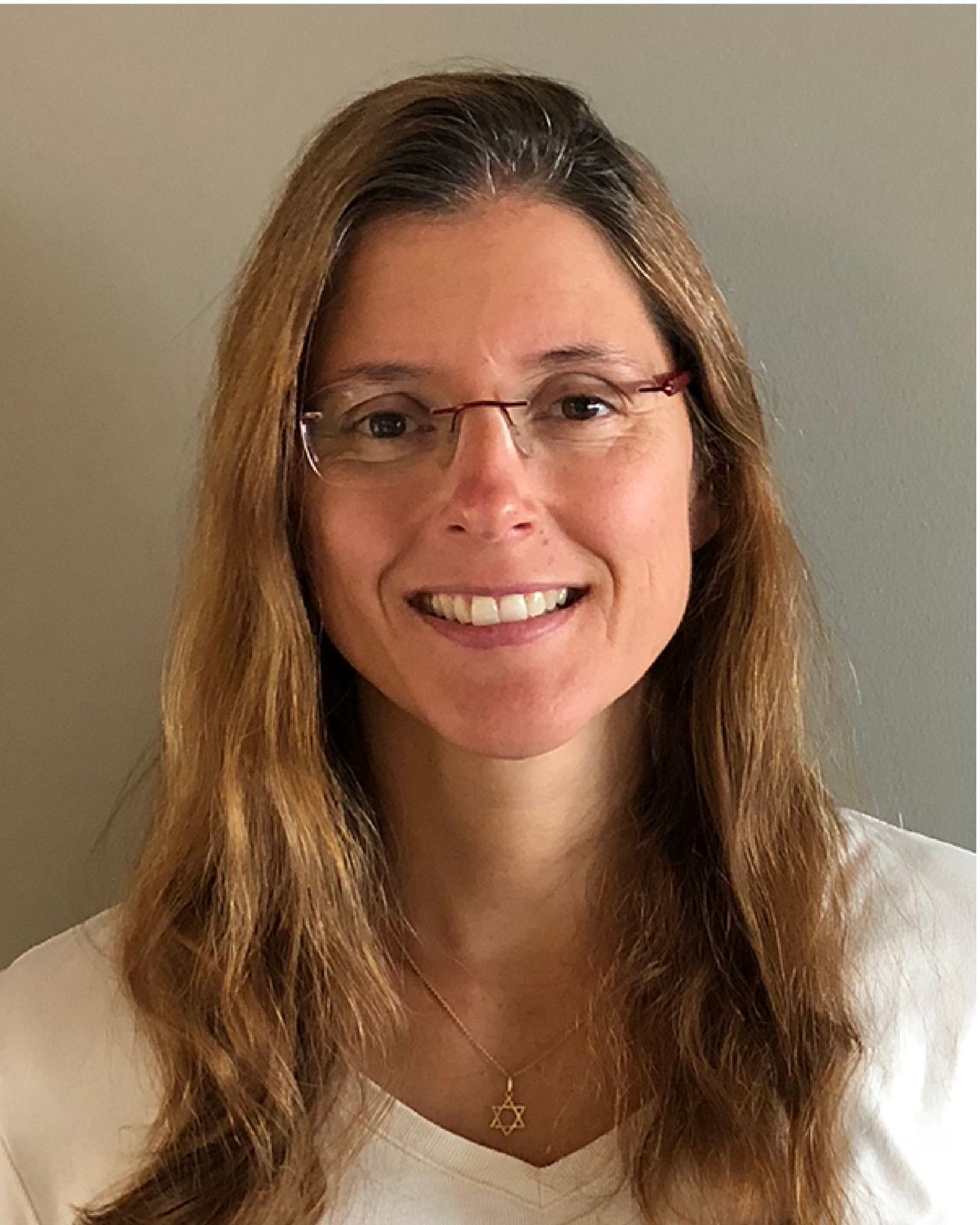}}]{Elisa R. Heymann}
is a Senior Scientist at the NSF Cybersecurity Center of Excellence at the
University of Wisconsin-Madison and an Associate Professor at the Autonomous
University of Barcelona. She was also in charge of the Grid/Cloud security
group at the UAB and participated in major Grid European Projects.
Dr. Heymann's research interests include software security and resource
management for Grid and Cloud environments. Her research is supported by
the NSF, Spanish government, the European Commission, and NATO.
\end{IEEEbiography}






\end{document}